 \documentclass[final,3p,times,authoryear]{elsarticle}

\usepackage{amssymb}
\usepackage{amsmath}

\journal{Nuclear Physics B}

\begin{document}

\begin{frontmatter}



 \author[1]{Vinesh Vijayan\corref{cor1}}
 \ead{Vinesh.physics@rathinam.in}
\affiliation[1]{organization={Department of Science \& Humanities},
		addressline={Rathinam Technical Campus}, 
		city={Coimbatore},
		postcode={641021}, 
		state={Tamilnadu},
		country={India}}
 \author[2]{Pasupuleti Thejasree} 
\affiliation[2]{organization={Department of Mechanical Engineering , Mohan Babu University,},
   		city={Tirupati},
   		postcode={517102}, 
   		state={Andhra Pradesh},
   		country={India}}
 \author[1]{P Satish Kumar }
  \author[1]{K Suganya}

\title{Geometric Formulation of Combined Conservative–Dissipative Mechanics via Contact Hamiltonian Dynamics: Symmetries, Reduction, and Variational Integrators} 




\begin{abstract}
We develop a unified geometric framework for mechanical systems that combine conservative and dissipative dynamics by formulating them on contact manifolds. Within this setting, we identify the Reeb vector field as the intrinsic generator of irreversibility and derive explicit laws describing how dissipation modifies symmetry reduction and momentum evolution. As a concrete application, we construct the contact Hamiltonian formulation of the rigid body with isotropic and anisotropic damping, classify all equilibrium configurations, and analyze their stability. Building on this continuous formulation, we design a second-order structure-preserving contact variational integrator obtained by a symmetric splitting of kinetic, potential, and dissipative components. Numerical experiments for representative dissipative systems demonstrate accurate energy decay, geometric consistency, and recovery of the symplectic Verlet scheme in the conservative limit. The proposed framework provides a coherent connection between the geometry of contact dynamics, physical irreversibility, and numerically stable integration, offering new tools for the analysis of mixed conservative–dissipative mechanical systems.
\end{abstract}




\end{frontmatter}



\section{Introduction}
\label{sec1}
Classical mechanics is traditionally formulated on symplectic manifolds, where Hamilton’s principle ensures conservation of energy, momentum, and the symplectic structure \cite{Weinstein1971, MarsdenWeinstein1974, Weinstein1978, Souriau1997}. Many physical systems, however, exhibit dissipation. Standard Lagrangian and Hamiltonian formalisms require modification to incorporate dissipative effects, which typically destroy conserved quantities and break symplectic invariance. \emph{Contact Hamiltonian mechanics}, a close analogue of symplectic geometry adapted to non-conservative systems, provides a natural geometric framework for modeling dissipation. Beginning with the work of Bravetti and others, contact geometry has emerged as a powerful tool for formulating friction, damping, thermodynamics, and nonholonomic phenomena in a coordinate-free manner \cite{Herglotz1930, Bravetti2016, Bravetti2017Survey, DundarAyar2021, MacielOrtizSchaerer2023, deLeon2019, deLeonLainz2019, OhsawaBloch2011, Galley2013}.

The foundational role of contact geometry as a framework for dissipation and thermodynamic formalism is established in \cite{MrugalaNultonSchonSalamon1990, Grmela2014, BravettiTapias2016}, where canonical equations and their physical motivations are developed. Although the symplectic literature has produced a general theory of momentum maps and contact reduction, these contributions remain largely abstract and are supported by only a limited number of numerical examples \cite{Gaset2023, GarciaMauriino2022}. Discrete Herglotz-based integrators---together with rigorous backward error analysis---are presented in the seminal works \cite{VermeerenBravettiSeri2019, Esen2024, MacielOrtizSchaerer2023, Cannarsa2020_English, Tapias2016Integrator}, addressing the preservation of geometric structure within the discrete Herglotz variational principle. However, the detailed interplay between symmetry and reduction in this context has yet to be explored. While the contact formalism has been extended to Lie algebroids and field theories, yielding elegant and highly general frameworks, these developments remain predominantly theoretical and lack concrete computational illustrations \cite{AnahorySimoes2025_English, deLeon2022HJ}.

Overall, the existing literature offers robust tools---such as the contact formalism, contact momentum maps, discrete Herglotz integrators, and Lie-algebroid generalizations---but it still lacks: (i) a fully worked example of a mechanically significant system with symmetries; (ii) a classification of isotropic or singular damping equilibria; (iii) the construction and testing of a discrete contact integrator for a nontrivial system; and (iv) an accessible connection between momentum-map reduction and its physical interpretation. This work fills these gaps by:

\begin{enumerate}
\item presenting a hybrid geometric--physical derivation of contact Hamiltonian mechanics for dissipative systems;
\item deriving general symmetry and momentum-map evolution equations that make explicit how dissipation breaks conservation laws;
\item providing a detailed analysis of the dissipative rigid body on $SO(3)$, including the classification of equilibria under anisotropic and singular damping; and
\item constructing a second-order contact variational integrator and applying it to a two-degree-of-freedom damped particle.
\end{enumerate}

Our aim is to develop a coherent framework that is both geometrically rigorous and physically accessible, while delivering new analytic and numerical results.
\section{Contact Hamiltonian Mechanics}\label{sec2}
\subsection{Contact manifolds and the canonical structure}\label{subsec1}
A \emph{contact manifold} is an odd-dimensional smooth manifold endowed with a maximally non-integrable geometric structure. Formally, it is a pair \((\mathcal{M}^{2n+1}, \zeta)\), where \(\zeta\) is a hyperplane distribution that may be expressed as the kernel of a 1-form \(\alpha\):
\begin{equation}\label{eq:contact-distribution}
    \zeta = \ker(\alpha).
\end{equation}
The 1-form \(\alpha\) satisfies the \emph{contact condition}
\begin{equation}\label{eq:contact-condition}
    \alpha \wedge (d\alpha)^n \neq 0 
    \qquad \text{everywhere on } \mathcal{M},
\end{equation}
which guarantees that \(\alpha \wedge (d\alpha)^n\) defines a volume form and that the distribution \(\zeta\) is maximally non-integrable.

\paragraph{Notation and convention.}
In what follows, we adopt the convention commonly used in the contact mechanics literature (e.g.\ Bravetti, de~León):

- \(\alpha\) denotes a \emph{general contact form} defining the distribution \(\zeta\) on an arbitrary contact manifold;
- \(\eta\) denotes the \emph{canonical contact form} in the model space \(T^*Q \times \mathbb{R}\), typically written in canonical coordinates \((q^i, p_i, s)\) as
\begin{equation}\label{eq:canonical-contact-form}
    \eta = ds - p_i\,dq^i .
\end{equation}

With this convention, all general definitions (e.g.\ the Reeb vector field, contact Hamiltonian vector fields, and abstract constructions on arbitrary contact manifolds) will employ the symbol \(\alpha\). The symbol \(\eta\) will be used exclusively in formulas expressed in canonical coordinates on \(T^*Q \times \mathbb{R}\).
This ensures consistency across general geometric definitions and canonical-coordinate expressions.
\subsection{Reeb vector field}\label{subsec2}
Given a contact manifold \((\mathcal{M}^{2n+1},\alpha)\), the \emph{Reeb vector field} \(R\)  \cite{Reeb1952_English} is the uniquely 
defined vector field characterized by
\begin{equation}\label{eq:ReebDef}
    \alpha(R)=1, 
    \qquad 
    \iota_R d\alpha = 0 .
\end{equation}
The condition \(\alpha(R)=1\) ensures that \(R\) is everywhere transverse to the contact distribution \(\zeta = \ker(\alpha)\), while the condition \(\iota_R d\alpha = 0\) requires that \(R\) spans the unique null direction of the two-form \(d\alpha\) when restricted to 
\(\zeta\). Existence and uniqueness follow from the contact condition

\begin{equation}\label{eq:contact-condition-Reeb}
    \alpha \wedge (d\alpha)^n \neq 0 ,
\end{equation}
which guarantees that \(d\alpha\) is maximally non-degenerate on \(\zeta\) but necessarily degenerate in exactly one transverse direction, singled out by \(R\).

\paragraph{Canonical coordinates.}
For the canonical contact manifold
\begin{equation}\label{eq:canonical-model}
    \mathcal{M} = T^*Q \times \mathbb{R}_s,
    \qquad 
    \eta = ds - p_i\,dq^i ,
\end{equation}
a general vector field can be expressed as
\begin{equation}\label{eq:general-vector-field}
    X 
    = 
    a_i\,\partial_{q^i}
    + b_i\,\partial_{p_i}
    + c\,\partial_s .
\end{equation}
Substituting \eqref{eq:general-vector-field} into the defining relations 
\eqref{eq:ReebDef} (with \(\alpha = \eta\)) yields
\[
    a_i = 0, 
    \qquad 
    b_i = 0,
    \qquad 
    c = 1,
\]
and therefore the Reeb vector field in canonical coordinates is
\begin{equation}\label{eq:Reeb-in-coordinates}
    R = \partial_s .
\end{equation}

\paragraph{Geometric interpretation.}
The Reeb vector field identifies the distinguished transverse direction compatible with the twisting determined by \(d\alpha\). Its flow preserves the contact form,
\begin{equation}\label{eq:Reeb-preserves-contact-form}
    \mathcal{L}_R \alpha = 0,
\end{equation}
and therefore preserves the contact structure itself. In contact Hamiltonian dynamics, the component of the dynamics along \(R\) encodes the intrinsic non-conservative behavior. For a contact Hamiltonian \(H\colon \mathcal{M} \to \mathbb{R}\), one has the evolution identity
\begin{equation}\label{eq:H-evolution-Reeb}
    \dot{H} = -\,R(H),
\end{equation}
which highlights the role of \(R\) as a dissipation or relaxation direction. In this sense Reeb field is the generator of irreversibility in contact Hamiltonian Mechanics.
\begin{figure}[t]
    \centering
    \includegraphics[width=0.45\linewidth]{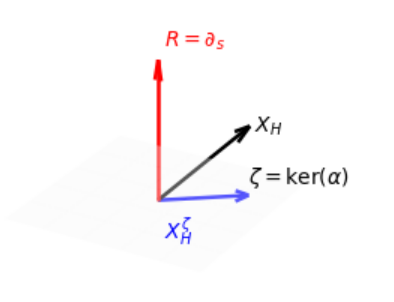}
    \caption{ Geometric structure of contact Hamiltonian dynamics. The plane represents the contact distribution $\zeta=\ker(\alpha)$, while the vertical arrow is the Reeb vector field $R=\partial_s$, which generates the non-conservative (dissipative) component of the dynamics. The blue arrow shows the tangential component $X_H^{\zeta}$ of the contact Hamiltonian vector field, lying entirely within $\zeta$, and the tilted black arrow depicts the full contact vector field $X_H$, whose vertical component encodes the irreversible contribution $R(H)$. This schematic illustrates the decomposition of contact dynamics into conservative and dissipative directions.
    }
    \label{fig1}
\end{figure}
Reeb dynamics plays a central role in contact topology, appearing in the study of closed Reeb orbits, the Weinstein conjecture, Legendrian geometry, and the analytical framework of symplectic field theory.

Figure\ref{fig1} schematically illustrates the decomposition of the contact Hamiltonian vector field into tangential (conservative) and vertical (dissipative) components on the contact manifold. The Reeb vector field identifies the direction of irreversibility, emphasizing the geometric origin of irreversible behavior.

\subsection{Contact Hamiltonian vector field}

Given a contact manifold \((\mathcal{M}^{2n+1},\alpha)\) with Reeb vector field \(R\), a smooth function 
\(H : \mathcal{M} \to \mathbb{R}\) defines a \emph{contact Hamiltonian vector field} \(X_H\) through the relations
\begin{equation}\label{eq:contact-vector-field}
    \iota_{X_H} d\alpha 
        = dH - (R(H))\,\alpha,
    \qquad
    \iota_{X_H}\alpha = -H .
\end{equation}
These equations uniquely determine \(X_H\).

\paragraph{Canonical coordinates.}
For the canonical contact manifold
\begin{equation}\label{eq:canonical-model-ham}
   \mathcal{M} = T^*Q \times \mathbb{R}_s,
    \qquad 
    \eta = ds - p_i\,dq^i ,
\end{equation}
one has
\begin{equation}\label{eq:canonical-deta}
    d\eta = -\,dp_i \wedge dq^i ,
\end{equation}
and the associated Reeb vector field is
\begin{equation}\label{eq:Reeb-canonical}
    R = \partial_s .
\end{equation}

A general vector field in canonical coordinates is written as
\begin{equation}\label{eq:XH-general}
    X_H
    = \dot q^i\,\partial_{q^i}
    + \dot p_i\,\partial_{p_i}
    + \dot s\,\partial_s .
\end{equation}
Here $s$ is the minimal additional variable—called the contact coordinate—needed to embed general non-conservative mechanical systems into a Hamiltonian-like framework while retaining geometric structure.
The contact form \ref{eq:canonical-deta} ensures that the combined mechanical–dissipative dynamics remains a first-order geometric flow, and its evolution is governed by \ref{eq:Reeb-canonical}, through which the dissipation law enters the equations of motion.
Because dissipative systems depend on the history of the dynamics, the contact coordinate $s$ records this history by accumulating the action or energy dissipated along the trajectory and thereby allows non-conservative systems to be written in Hamiltonian form.

Using \(\iota_{X_H}\eta = -H\), we obtain
\begin{equation}\label{eq:temp-s-eqn}
    \dot{s} - p_i\,\dot{q}^i = -\,H .
\end{equation}

Next, using \eqref{eq:canonical-deta}, we compute
\begin{equation}\label{eq:iXHdeta}
    \iota_{X_H} d\eta
    = \dot q^i\,dp_i - \dot p_i\,dq^i .
\end{equation}
On the other hand,
\begin{equation}\label{eq:dH}
    dH
    =
    \frac{\partial H}{\partial q^i}\,dq^i
    + \frac{\partial H}{\partial p_i}\,dp_i
    + \frac{\partial H}{\partial s}\,ds,
    \qquad
    R(H) = \partial_s H ,
\end{equation}
so
\begin{equation}\label{eq:dH-minus-RHeta}
    dH - R(H)\,\eta
    =
    \left(
        \frac{\partial H}{\partial q^i}
        + p_i\frac{\partial H}{\partial s}
    \right) dq^i
    +
    \frac{\partial H}{\partial p_i}\,dp_i .
\end{equation}

Matching coefficients of \(dq^i\) and \(dp_i\) between 
\eqref{eq:iXHdeta} and \eqref{eq:dH-minus-RHeta} yields the evolution equations:
\begin{equation}\label{eq:qdot}
    \dot{q}^i 
    = \frac{\partial H}{\partial p_i},
\end{equation}
\begin{equation}\label{eq:pdot}
    \dot{p}_i
    = -\frac{\partial H}{\partial q^i}
      - p_i\frac{\partial H}{\partial s}.
\end{equation}

Substituting \eqref{eq:qdot} into \eqref{eq:temp-s-eqn} gives the final component:
\begin{equation}\label{eq:sdot}
    \dot{s}
    = p_i\frac{\partial H}{\partial p_i} - H .
\end{equation}

Thus, the \emph{contact Hamilton equations} in canonical coordinates are
\begin{equation}\label{eq:canonical-contact-equations}
    \dot q^i 
        = \frac{\partial H}{\partial p_i},
    \qquad
    \dot p_i 
        = -\frac{\partial H}{\partial q^i}
          - p_i\frac{\partial H}{\partial s},
    \qquad
    \dot s 
        = p_i\frac{\partial H}{\partial p_i} - H .
\end{equation}

If \(H\) is independent of \(s\), then \(\partial_s H = 0\), and the subsystem in \((q^i,p_i)\) reduces to the usual Hamilton equations, while
\begin{equation}\label{eq:sdot-s-independent}
    \dot{s} 
    = p_i\frac{\partial H}{\partial p_i} - H .
\end{equation}

\subsection{Mechanical--dissipative decomposition and energy balance}

We now consider a contact Hamiltonian of separated form
\begin{equation}\label{eq:H-separated}
    H(q,p,s) = H_0(q,p) + \Gamma(s),
\end{equation}
where \(H_0\) is the mechanical (conservative) Hamiltonian and \(\Gamma(s)\) encodes the dissipative contribution.  In accordance with our convention, dissipation enters through
\begin{equation}\label{eq:Gamma-prime-lambda}
    \Gamma'(s) = \lambda ,
\end{equation}
so that \(\lambda\) acts as an effective dissipation parameter in the contact dynamics.

Because \(H_0\) depends only on \((q,p)\) and \(\Gamma\) only on \(s\), one has
\begin{equation}\label{eq:H-derivatives-separated}
    \frac{\partial H}{\partial p_i}
        = \frac{\partial H_0}{\partial p_i},
    \qquad
    \frac{\partial H}{\partial q^i}
        = \frac{\partial H_0}{\partial q^i},
    \qquad
    \frac{\partial H}{\partial s}
        = \Gamma'(s) = \lambda .
\end{equation}

Substituting \eqref{eq:H-derivatives-separated} into the contact Hamilton equations \eqref{eq:canonical-contact-equations}, we obtain the dynamical system

\begin{equation}\label{eq:contact-separated-system}
    \dot{q}^{\,i}
        = \frac{\partial H_0}{\partial p_i},
    \qquad
    \dot{p}_i
        = -\,\frac{\partial H_0}{\partial q^i}
          - \lambda\,p_i,
    \qquad
    \dot{s}
        = p_i\frac{\partial H_0}{\partial p_i}
          - H_0 - \Gamma(s).
\end{equation}

The term
\[
    -\,\lambda\,p_i
\]
represents a Rayleigh-type friction force acting on the momentum variables. Thus \(\lambda\) appears as the effective damping coefficient generated by the contact Hamiltonian structure.

The evolution of \(s\) encodes the balance between mechanical power input \(p_i \frac{\partial H_0}{\partial p_i}\) and the total Hamiltonian \(H = H_0 + \Gamma(s)\), providing a geometric formulation of energy dissipation within the contact framework.

From this point onward, we assume \(Q = \mathbb{R}^n\) equipped with the Euclidean metric, so that
\begin{equation}\label{eq:euclidean-metric}
    g_{ij} = \delta_{ij},
    \qquad
    g^{ij} = \delta^{ij},
\end{equation}
and we freely use vector notation such as 
\(\|p\|^2\) and \(\nabla V\)
when convenient.

\subsubsection{Quadratic mechanical Hamiltonian}

If the mechanical energy has the standard form
\begin{equation}\label{eq:H0-quadratic}
    H_0(q,p)
      = \frac{1}{2}\, g^{ij}(q)\,p_i p_j + V(q),
\end{equation}
then the momentum--velocity relation follows immediately:
\begin{equation}\label{eq:mom-vel-relation}
    \frac{\partial H_0}{\partial p_i}
      = g^{ij}(q)\,p_j
      = \dot{q}^{\,i}.
\end{equation}
The momentum equation becomes
\begin{equation}\label{eq:pdot-quadratic}
    \dot{p}_i
     = -\,\frac{\partial V}{\partial q^i}
       - \Gamma'(s)\,p_i .
\end{equation}

Expressing the dynamics in configuration variables yields the damped Newton equation
\begin{equation}\label{eq:damped-Newton}
    g_{ij}(q)\,\ddot{q}^{\,j}
    + \Gamma'(s)\,g_{ij}(q)\,\dot{q}^{\,j}
    + \frac{\partial V}{\partial q^i}
    = 0 .
\end{equation}

\subsubsection*{Energy dissipation}

To compute the time derivative of the mechanical energy \(H_0\), differentiate:
\begin{equation}\label{eq:dH0dt-general}
    \frac{dH_0}{dt}
      = \frac{\partial H_0}{\partial q^i}\,\dot{q}^{\,i}
        + \frac{\partial H_0}{\partial p_i}\,\dot{p}_i .
\end{equation}
Using the equations of motion, the conservative terms cancel, yielding the intrinsic
contact dissipation relation
\begin{equation}\label{eq:dH0dt-intrinsic}
    \frac{dH_0}{dt}
      = -\,\Gamma'(s)\,
         p_i\,\frac{\partial H_0}{\partial p_i}.
\end{equation}

\begin{figure}[t]
    \centering
    \includegraphics[width=0.35\linewidth]{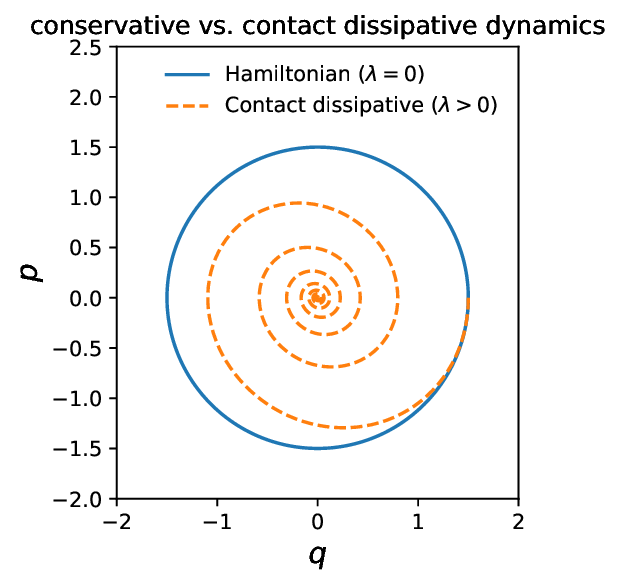}
    \caption{Phase portrait of the harmonic oscillator illustrating the difference between conservative Hamiltonian dynamics ($\lambda = 0$) and contact dissipative dynamics     with linear dissipation $\Gamma'(s)=\lambda>0$. The solid curves correspond to the Hamiltonian flow, which preserves the level sets of the mechanical nergy$H_0=\tfrac12(p^2+q^2)$ and therefore yields closed orbits in the  $(q,p)$-plane.  The dashed curves correspond to the contact Hamiltonian flow, for which the momentum  evolution $p\dot{} = -q - \lambda p$ induces exponential decay of mechanical energy.  As predicted by the contact Hamiltonian equations, trajectories spiral toward the origin, demonstrating the geometric effect of dissipation generated by the Reeb component $R(H)=\lambda$. 
    }
    \label{fig2}
\end{figure}

For the quadratic kinetic energy \eqref{eq:H0-quadratic},
\begin{equation}\label{eq:p-dHdpi}
    p_i\,\frac{\partial H_0}{\partial p_i}
      = g^{ij}(q)\,p_i p_j ,
\end{equation}
so
\begin{equation}\label{eq:dH0dt-quadratic}
    \frac{dH_0}{dt}
      = -\,\Gamma'(s)\,
        g^{ij}(q)\,p_i p_j
      \le 0
      \qquad (\Gamma'(s) \ge 0).
\end{equation}
Thus the mechanical energy decreases monotonically whenever the dissipation function satisfies \(\Gamma'(s)\ge 0\). Figure \ref{fig2} illustrates the geometric energy decay induced by the Reeb vector field contribution for a harmonic oscillator. In the conservative case ($\lambda=0$), the energy orbits are closed, whereas with dissipation ($\lambda \neq =0$), the contact flow generates spirals that converge to the origin.

\subsubsection*{Linear dissipation}

For a linear dissipation potential,
\begin{equation}\label{eq:Gamma-linear}
    \Gamma(s) = \lambda s,
    \qquad \lambda > 0,
\end{equation}
we have the constant derivative
\begin{equation}\label{eq:Gamma-prime-linear}
    \Gamma'(s) = \lambda.
\end{equation}
Then
\begin{equation}\label{eq:dH0dt-linear}
    \frac{dH_0}{dt}
      = -\,\lambda\,g^{ij}(q)\,p_i p_j,
\end{equation}
the familiar expression for linear (viscous) mechanical damping.

Hence contact Hamiltonian mechanics naturally reproduces standard dissipative dynamics while maintaining a geometric Hamiltonian structure.
\section{Symmetry and Momentum Maps in Contact Mechanics}

\subsection{Group actions and cotangent lifts}

Let \(G\) be a Lie group acting smoothly on the configuration space \(Q\),
\begin{equation}\label{eq:G-action-Q}
    \Phi : G \times Q \to Q, 
    \qquad (g,q) \mapsto g\cdot q .
\end{equation}
This action induces a cotangent-lifted action on \(T^{*}Q\) preserving the canonical one-form \(\theta = p_i\,dq^i\). Extending the action trivially in the \(s\)-coordinate yields a strict contact action on
\begin{equation}\label{eq:M-contact-space}
   \mathcal{ M} = T^{*}Q \times \mathbb{R}_s
\end{equation}
via
\begin{equation}\label{eq:contact-lift-action}
    \tilde{\Phi}(g)(q,p,s)
      = \bigl( g\cdot q , (T^{*}g)^{-1}p , s \bigr).
\end{equation}
Since cotangent lifts preserve \(\theta\) and leave \(s\) unchanged, the
canonical contact form
\begin{equation}\label{eq:contact-form}
    \eta = ds - p_i\,dq^i
\end{equation}
is invariant:
\begin{equation}\label{eq:strict-invariance}
    \tilde{\Phi}(g)^{*}\eta = \eta .
\end{equation}
Thus the action proceeds by \emph{strict contactomorphisms}.

In general, a contact action satisfies the weaker conformal relation
\begin{equation}\label{eq:conformal-contact}
    \Phi_g^*\eta = e^{\sigma(g)} \eta ,
\end{equation}
in which case the associated momentum map includes a correction term depending
on \(\sigma(g)\).  
In this work we restrict attention to strictly invariant actions:
\begin{equation}\label{eq:strict-action}
    \Phi_g^*\eta = \eta ,
\end{equation}
a condition automatically satisfied by cotangent lifts extended trivially in \(s\).

\paragraph{Infinitesimal generators and the classical momentum map.} For each \(\xi \in \mathfrak{g}\), denote its infinitesimal generator on \(Q\)
by \(\xi_Q\), and its cotangent lift on \(T^{*}Q\) by \(\xi_{T^{*}Q}\).  
The associated momentum map
\(J: T^{*}Q \to \mathfrak{g}^{*}\) is defined by
\begin{equation}\label{eq:cotangent-momentum}
    \langle J(q,p), \xi \rangle 
        = p\!\left( \xi_Q(q) \right).
\end{equation}
Since the \(s\)-coordinate plays no role in \eqref{eq:cotangent-momentum}, the same formula defines a momentum map on \(\mathcal{M} = T^{*}Q \times \mathbb{R}\).

Let \(\xi_M\) denote the infinitesimal generator of the strict contact action
on \(M\).  
Strictness implies
\begin{equation}\label{eq:xiM-Lie-eta}
    \mathcal{L}_{\xi_M}\eta = - dJ_\xi,
    \qquad 
    \mathcal{L}_{\xi_M} d\eta = dJ_\xi .
\end{equation}

\paragraph{Derivative of the momentum map.}
Along the contact Hamiltonian vector field \(X_H\),
\begin{equation}\label{eq:Jdot-def}
    \dot{J}_\xi 
        = X_H(J_\xi) 
        = dJ_\xi(X_H)
        = \mathcal{L}_{X_H} J_\xi .
\end{equation}
Using \eqref{eq:xiM-Lie-eta}, we compute
\begin{equation}\label{eq:momentum-evolution-step}
    \mathcal{L}_{X_H} dJ_\xi
      = \mathcal{L}_{X_H} \mathcal{L}_{\xi_M} d\eta
      = - \mathcal{L}_{\xi_M} \mathcal{L}_{X_H} d\eta .
\end{equation}
The contact Hamiltonian structure gives
\begin{equation}\label{eq:LXH-deta}
    \mathcal{L}_{X_H} d\eta = dH - (R(H))\,\eta .
\end{equation}
Substituting \eqref{eq:LXH-deta} into \eqref{eq:momentum-evolution-step} yields
\begin{equation}\label{eq:Jdot-intermediate}
\begin{aligned}
    \dot{J}_\xi
        &= - \mathcal{L}_{\xi_M}( dH - (R(H))\,\eta ) \\
        &= -\,\xi_M(H)
           + \bigl(\mathcal{L}_{\xi_M}(R(H))\bigr)\,\eta .
\end{aligned}
\end{equation}

\subsection{Evolution of the momentum map under contact dynamics}

Assume that the contact Hamiltonian \(H:\mathcal{M}\to \mathbb{R}\) is \(G\)-invariant:
\begin{equation}\label{eq:H-G-invariant}
    H(\tilde{\Phi}(g)(m)) = H(m)
    \qquad \forall g\in G,\; m\in M.
\end{equation}
Then \(\xi_M(H) = 0\) for all \(\xi\), so \eqref{eq:Jdot-intermediate} reduces to
\begin{equation}\label{eq:momentum-decay-general}
    \dot{J}_\xi 
      = - (R(H))\,J_\xi .
\end{equation}

For the separated Hamiltonian
\begin{equation}\label{eq:separated-H}
    H(q,p,s) = H_0(q,p) + \Gamma(s),
\end{equation}
one has
\begin{equation}\label{eq:R(H)-lambda}
    R(H) = \Gamma'(s) = \lambda,
\end{equation}
and therefore
\begin{equation}\label{eq:Jxi-decay-lambda}
    \dot{J}_\xi = -\lambda\,J_\xi .
\end{equation}
Integrating,
\begin{equation}\label{eq:J-decay-solution}
    J_\xi(t)
      = e^{-\lambda t}\, J_\xi(0).
\end{equation}

\begin{figure}[t]
    \centering
    \includegraphics[width=0.45\linewidth]{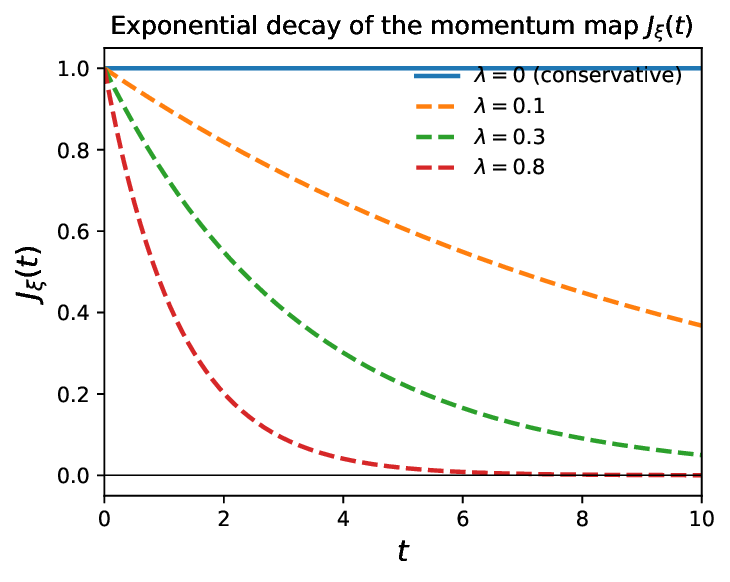}
    \caption{ Exponential decay of the momentum map $J_{\xi}(t)$ under contact Hamiltonian dynamics.     For a strictly invariant group action and a separated contact Hamiltonian     $H(q,p,s)=H_{0}(q,p)+\Gamma(s)$, the evolution of the momentum map satisfies     $\dot{J}_{\xi} = -\,\Gamma'(s)\,J_{\xi}$, yielding the explicit solution     $J_{\xi}(t)=J_{\xi}(0)e^{-\lambda t}$ with $\lambda=\Gamma'(s)$. The solid curve ($\lambda=0$) represents the conservative Hamiltonian case, for which 
    momenta are conserved.  The dashed curves ($\lambda>0$) illustrate the strictly dissipative contact case, where the Reeb contribution $R(H)=\lambda$ drives exponential decay of all momentum components, reflecting the loss of symmetry-induced conservation laws.
    }
    \label{fig3}
\end{figure}

Thus, unlike in symplectic Hamiltonian mechanics where symmetry implies conservation, here symmetry only guarantees \emph{exponential decay} of the
momentum map whenever \(\lambda \neq 0\).

Figure \ref{fig3} shows the time evolution of the momentum map under contact dynamics. The solid curve represents exact conservation, while the dashed curves illustrate exponential decay governed by equation (\ref{eq:J-decay-solution}), for various values of $\lambda$, indicating symmetry breaking due to dissipation.

\subsection{Obstruction to symplectic reduction}

In symplectic mechanics, the Marsden--Weinstein reduction requires that each momentum level set
\begin{equation}\label{eq:level-set}
    J^{-1}(\mu)
\end{equation}
is invariant under the Hamiltonian flow.  
Equation \eqref{eq:Jxi-decay-lambda} implies
\begin{equation}\label{eq:level-set-not-invariant}
    \frac{d}{dt} J_\xi = -\lambda\,J_\xi ,
\end{equation}
so trajectories satisfy
\begin{equation}\label{eq:not-constant-momentum}
    J(m(t)) \neq \mu 
    \qquad \text{for } t>0
\end{equation}
unless \(\lambda = 0\) or \(J_\xi = 0\).  
This violates the key requirement for symplectic reduction.

Nevertheless, reduction can be adapted to contact dynamics in two natural ways:

\begin{itemize}

    \item \emph{Time-dependent reduction.}  
    For constant \(\lambda\), one may reduce along the exponentially decaying
    momentum level
    \begin{equation}\label{eq:time-dependent-level}
        \mu(t) = e^{-\lambda t}\,\mu_0 ,
    \end{equation}
    which tracks the exact decay of \(J_\xi(t)\).

    \item \emph{Reduction via symplectification.}  
    Embedding the contact manifold into its symplectization
    \begin{equation}\label{eq:symplectification}
        (\mathcal{M} \times \mathbb{R}_{+},\; \omega = d(r\eta)),
    \end{equation}
    allows one to apply classical symplectic reduction on the extended space and then descend to a reduced contact manifold.

\end{itemize}

These approaches show that although dissipation obstructs standard symplectic reduction, suitable modifications restore a meaningful reduction procedure in the contact setting.
\section{Rigid Body with Dissipation on $SO(3)$}

In the rigid-body section, we will switch from index notation to boldface vector notation \((\mathbf{M},\boldsymbol{\Omega})\), using the Euclidean
identification of \(\mathfrak{so}(3)^*\) with \(\mathbb{R}^3\). All dot and cross products there will refer to this identification.
\subsection{Configuration and Contact Hamiltonian}

We consider a rigid body whose configuration space is \(Q = SO(3)\).
Using left trivialization, elements of \(T^{*}SO(3)\) are represented as
\[
(R,M) \in SO(3)\times\mathbb{R}^3 ,
\]
where \(M \in \mathfrak{so}(3)^* \simeq \mathbb{R}^3\) is the body angular momentum.
Let
\[
\Omega = I^{-1}M
\]
denote the body angular velocity, where \(I\) is the inertia tensor.

To incorporate dissipation, we work on the contact manifold
\[
T^*SO(3)\times\mathbb{R}_s
\]
with contact Hamiltonian
\begin{equation}\label{eq:RB-H}
    H(R,M,s)
      = \tfrac12\, M\cdot I^{-1}M \;+\; \gamma s ,
      \qquad \gamma > 0 .
\end{equation}
The term \(\gamma s\) generates linear dissipation through the contact structure.

\subsection{Contact rigid-body equations}

Applying the contact Hamilton equations to \((R,M,s)\) yields
\begin{align}
\dot{R} &= R\,\widehat{\Omega}, 
    \qquad \Omega = I^{-1}M,
    \label{eq:rb-R}
\\[4pt]
\dot{M} &= M\times\Omega \;-\;\gamma M,
    \label{eq:rb-M}
\\[4pt]
\dot{s} &= \tfrac12\,M\cdot\Omega \;-\;\gamma s.
    \label{eq:rb-s}
\end{align}
Equation \eqref{eq:rb-R} is the standard rigid-body kinematics on \(SO(3)\). Equation \eqref{eq:rb-M} reproduces Euler’s equation with isotropic linear
damping of magnitude \(\gamma\). The \(s\)-equation describes the contact-induced dissipation channel.

\subsection{Symmetry and the momentum map}

The left action of \(SO(3)\) on \(T^*SO(3)\) has momentum map
\begin{equation}\label{eq:rb-J}
    J(R,M) = M .
\end{equation}
Under the contact flow generated by \eqref{eq:RB-H}, 
\begin{equation}\label{eq:rb-M-evolution}
    \dot{M} = M\times I^{-1}M \;-\;\gamma M,
\end{equation}
which coincides with \eqref{eq:rb-M}.
Thus the contact formulation recovers the coadjoint rigid-body dynamics together with the dissipative decay term induced by \(\partial_s H = \gamma\).
\begin{figure}[t]
    \centering
    \includegraphics[width=0.45\linewidth]{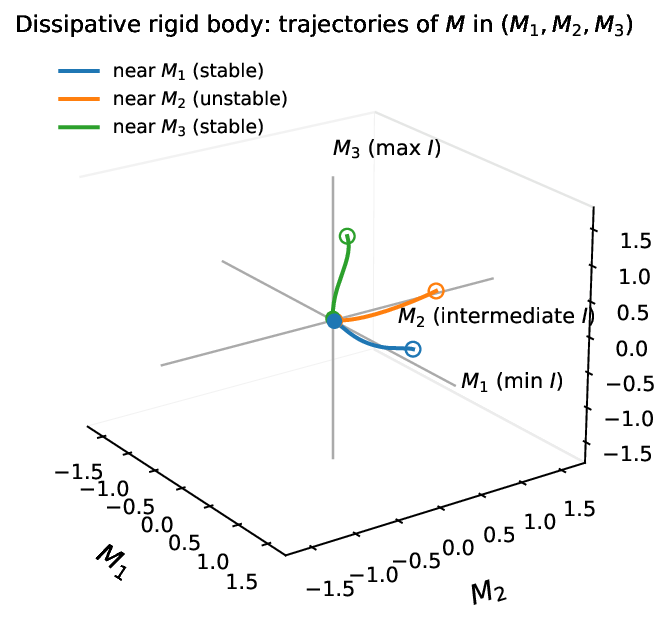}
    \caption{Trajectories of the dissipative rigid body dynamics $\dot M = M\times\Omega - \gamma M$ with $\Omega = I^{-1}M$ are shown in $(M_1,M_2,M_3)$-space for initial conditions near each principal axis. Solutions starting near the smallest and largest inertia axes remain close to those directions as they decay toward the origin, while trajectories near the intermediate axis peel away and migrate toward a stable axis, illustrating the classical stability structure of rigid-body rotation.  Dissipation drives all trajectories toward the origin, reflecting the monotone decay of the rotational kinetic energy $H_0 = \tfrac12 M\cdot I^{-1}M$ under the contact-type damping term $-\gamma M$.
    }
    \label{fig4}
\end{figure}

\subsection{Energy dissipation}

The mechanical energy
\begin{equation}\label{eq:H0-rb}
    H_0(M) = \tfrac12\,M\cdot I^{-1}M
\end{equation}
satisfies
\begin{equation}\label{eq:H0dot-rb}
\frac{d}{dt} H_0(M)
  = M\cdot I^{-1}\dot{M}
  = -2\gamma\,H_0(M),
\end{equation}
using \(M\cdot I^{-1}(M\times\Omega)=0\).
Hence
\begin{equation}\label{eq:H0-decay-rb}
    H_0(t) = H_0(0)\,e^{-2\gamma t},
\end{equation}
showing exponential decay of rotational kinetic energy.
Figure \ref{fig4} depicts phase space trajectories of a rigid body system in angular momentum space. Dissipation causes all trajectories to contract toward the origin, modifying the classical rigid-body stability structure with contact damping effects.
\subsection{Equilibria under anisotropic damping}

A more general model replaces the scalar \(\gamma\) with a positive-semidefinite
matrix \(D\):
\begin{equation}\label{eq:anisotropic}
    \dot{M} = M\times I^{-1}M \;-\; D M .
\end{equation}
An equilibrium \(M^{*}\) must satisfy
\begin{equation}\label{eq:eq-conditions}
    D M^{*} = 0,
    \qquad
    M^{*} \times I^{-1}M^{*} = 0 .
\end{equation}
The second condition means that \(M^{*}\) lies along a principal axis of inertia. The first requires \(M^{*} \in \ker D\).

\subsection{Stability of equilibria}

Linearizing \eqref{eq:anisotropic} about an equilibrium \(M^{*}\) gives
\begin{equation}\label{eq:linearized}
    \delta\dot{M}
      = \delta M \times \Omega^{*}
        + M^{*} \times I^{-1}\delta M
        - D\,\delta M ,
\end{equation}
with \(\Omega^{*} = I^{-1}M^{*}\).

The conservative part implies the classical rigid-body result:
\begin{itemize}
    \item rotations about the largest or smallest inertia axes are stable;
    \item rotation about the intermediate axis is a saddle-type unstable equilibrium.
\end{itemize}

The dissipative term \(-D\,\delta M\) causes all modes not in \(\ker D\) to decay exponentially.  Modes lying in \(\ker D\) behave as in the undamped rigid body.

\subsubsection{Stability Theorem}

Let \(M^{*}\) satisfy \eqref{eq:eq-conditions}. Then:

\begin{enumerate}
    \item If \(M^{*}\) lies on the maximal or minimal principal axes of inertia and belongs to \(\ker D\), then \(M^{*}\) is linearly asymptotically stable.

    \item If \(M^{*}\) lies on the intermediate principal axis and belongs to \(\ker D\), then \(M^{*}\) is a saddle-type unstable equilibrium.

    \item If \(M^{*} \notin \ker D\), then no nontrivial steady rotation exists and all trajectories decay to the origin.
\end{enumerate}

\paragraph{Theorem.}
The set of rigid-body equilibria under anisotropic damping is
\begin{equation}\label{eq:equilibrium-set}
\mathrm{Eq}
=
\{0\}
\;\cup\;
\{\,M\neq 0 : D M = 0,\ \ M \parallel \text{a principal axis of } I\,\}.
\end{equation}
Thus dissipation selects precisely those principal-axis rotations lying in
\(\ker D\); all other motions decay toward these sets.
\begin{figure}[t]
    \centering
    \includegraphics[width=0.45\linewidth]{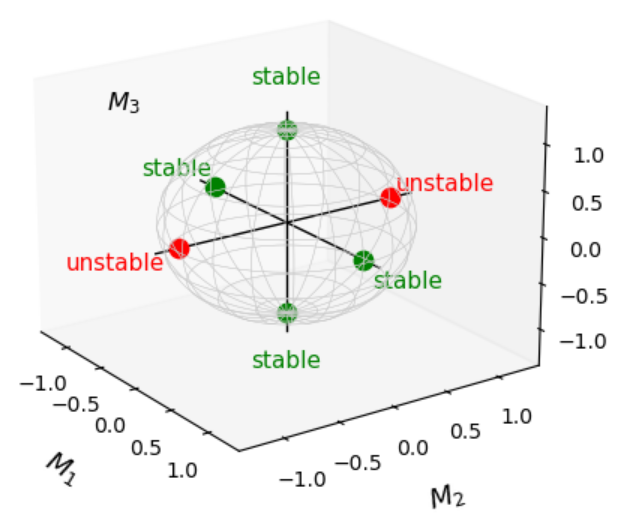}
    \caption{Equilibria of the dissipative rigid body $\dot M = M\times\Omega - \gamma M$ with $\Omega = I^{-1}M$ lie on the principal axes of inertia.  The smallest and largest inertia axes ($M_1$ and $M_3$) correspond to stable equilibria, while the intermediate axis ($M_2$) yields an unstable equilibrium, consistent with the classical rigid-body stability structure. The diagram shows stable (green) and unstable (red) equilibrium points on the momentum sphere, clearly illustrating the effect of the contact-type damping term on the qualitative dynamics.
    }
    \label{fig5}
\end{figure}

Figure \ref{fig5} visualizes stable (green) and unstable (red) equilibria on the momentum sphere. The stable configurations align with the largest and smallest inertia axes, while the intermediate axis remains unstable, consistent with the contact rigid body stability theorem.

\section{Contact Variational Integrators}

Contact mechanics generalizes Hamiltonian dynamics to systems with dissipation by introducing an additional action-like variable \(s\) and replacing Hamilton’s principle with the Herglotz variational principle.  
This viewpoint naturally suggests geometric numerical methods that respect the underlying contact structure.  
In this section we describe a practical and robust contact variational integrator based on Hamiltonian splitting, and we illustrate its accuracy on a two-degree-of-freedom damped particle.

\subsection{Discrete Herglotz principle}

In the continuous Herglotz formulation, curves \((q(t),s(t))\) satisfy
\begin{equation}\label{eq:cont-Herglotz}
    \dot{s} = L(q,\dot{q},s).
\end{equation}
The discrete analogue replaces the integral action by the recursion
\begin{equation}\label{eq:disc-Herglotz}
    s_{k+1} = s_k + h\,L(q_k,q_{k+1},s_k).
\end{equation}
Variations of \(\{q_k\}\) lead to the discrete contact Euler--Lagrange equations, which reduce to the classical discrete Euler--Lagrange equations when \(L\) is independent of \(s\).  

Although the discrete Herglotz principle yields rigorous variational integrators, the schemes are often implicit.  
To obtain an explicit, high-quality method, we instead adopt the contact Hamiltonian viewpoint, where splitting methods provide clean and effective structure-preserving discretizations.

\subsection{Hamiltonian splitting}

For mechanical systems of the form
\begin{equation}\label{eq:H-split}
    H(q,p,s) = T(p) + V(q) + \gamma s,
\end{equation}
the Hamiltonian separates naturally as
\begin{equation}\label{eq:ABC-Hamiltonians}
    H_A = T(p), \qquad 
    H_B = V(q), \qquad 
    H_C = \gamma s .
\end{equation}
Each part generates an \emph{exactly solvable} contact Hamiltonian flow.

\paragraph{$A$-flow (kinetic).}
\begin{equation}\label{eq:A-flow}
\begin{aligned}
    q &\mapsto q + h\,\partial_p T(p),\\
    p &\mapsto p,\\
    s &\mapsto s + h\,T(p).
\end{aligned}
\end{equation}

\paragraph{$B$-flow (potential).}
\begin{equation}\label{eq:B-flow}
\begin{aligned}
    q &\mapsto q,\\
    p &\mapsto p - h\,\nabla V(q),\\
    s &\mapsto s - h\,V(q).
\end{aligned}
\end{equation}

\paragraph{$C$-flow (damping).}
\begin{equation}\label{eq:C-flow}
\begin{aligned}
    q &\mapsto q,\\
    p &\mapsto e^{-\gamma h}p,\\
    s &\mapsto e^{-\gamma h}s.
\end{aligned}
\end{equation}

\begin{figure}[t]
    \centering
    \includegraphics[width=0.85\linewidth]{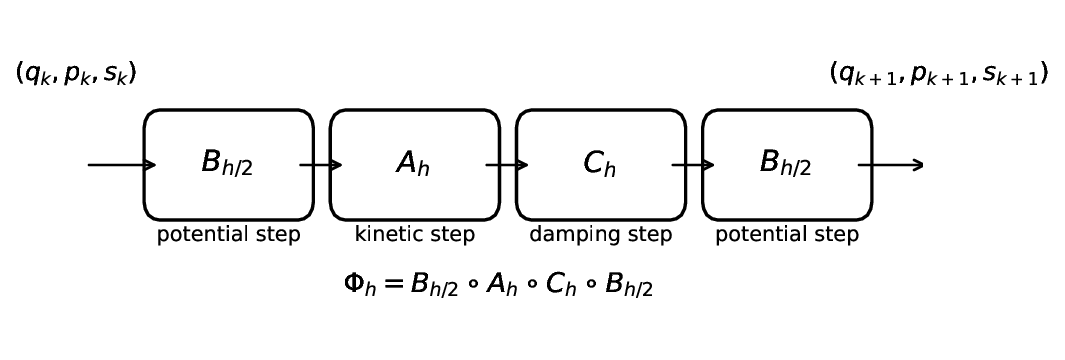}
    \caption{Flowchart representation of the contact splitting integrator $\Phi_h = B_{h/2}\circ A_h\circ C_h\circ B_{h/2}$ used to approximate the contact Hamiltonian dynamics. The maps $A_h$, $B_{h/2}$, and $C_h$ correspond respectively to the kinetic, potential, and damping subflows, each of which admits a closed-form solution. The symmetric Strang-type composition shown here forms the basis for the second-order accurate geometric integrator developed in Section~5.
    }
    \label{fig6}
\end{figure}

Composing these exact subflows yields a structure-preserving method. We use the symmetric Strang composition

\begin{equation}\label{eq:Strang-Integrator}
    \Phi_h 
    = B_{h/2} \circ A_h \circ C_h \circ B_{h/2}.
\end{equation}
Because each subflow is an exact contact transformation, the resulting integrator preserves the contact structure up to \(O(h^3)\), including the correct decay of both \(H_0\) and \(s\). Figure \ref{fig6} shows the flow chart of the variational integrator.

\subsection{Symplectic limit}

When dissipation vanishes, \(\gamma \to 0\),
\begin{equation}\label{eq:Ch-limit}
    C_h \to \mathrm{id}.
\end{equation}
The method reduces to
\begin{equation}\label{eq:VV-limit}
    \Phi_h \to B_{h/2} \circ A_h \circ B_{h/2},
\end{equation}
which is exactly the velocity-Verlet (Störmer--Verlet) integrator. Thus the contact variational integrator smoothly recovers symplectic Hamiltonian dynamics in the conservative limit.

\section{Numerical Example: 2-DOF Damped Particle}

\subsection{System}

We consider a particle in \(\mathbb{R}^2\) with Hamiltonian
\begin{equation}\label{eq:2DOF-H}
    H(q,p,s)
    = \tfrac12 \|p\|^2 + V(q) + \gamma s.
\end{equation}
The potential is
\begin{equation}\label{eq:2DOF-V}
    V(q_1,q_2)
    = \tfrac12(q_1^2 + q_2^2)
      + 0.1\,q_1^2 q_2 .
\end{equation}

\subsection{Integrator implementation}

We use the contact splitting integrator with time step
\begin{equation}\label{eq:step-size}
    h = 0.01.
\end{equation}
The simulation is run for total time \(T = 20\).  
For comparison, we also implement a classical RK4 method applied directly to 
the contact Hamiltonian equations.
\begin{figure}[t]
    \centering
    \includegraphics[width=\linewidth]{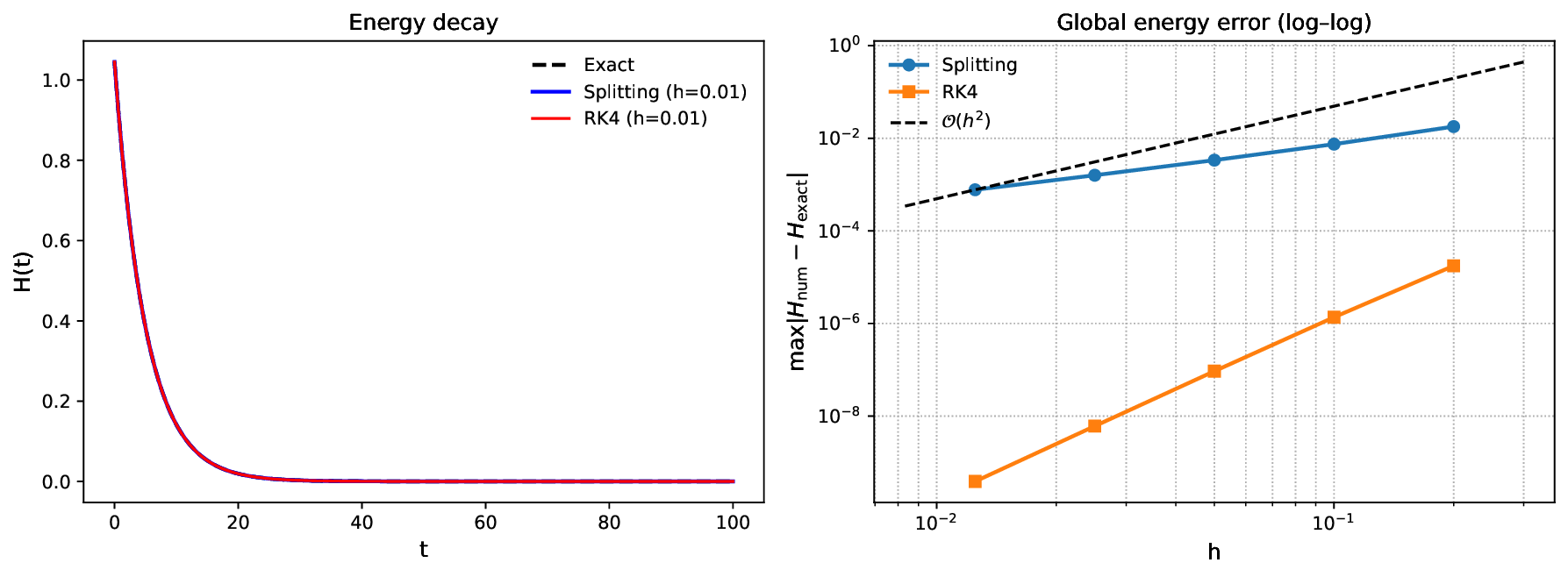}
    \caption{
        Left: energy decay for the contact Hamiltonian system, comparing the exact exponential law with the numerical solutions produced by the contact splitting method and by RK4 using the same step size $h=0.01$.  
    Right: log--log plot of the global energy error $\max_{t\in[0,T]}|H_{\mathrm{num}}(t)-H_{\mathrm{exact}}(t)|$ as a function of the step size, showing second-order convergence for both integrators and confirming the $\mathcal{O}(h^{2})$ behavior.  
    While RK4 exhibits a smaller error constant for this smooth test problem, only the geometric splitting method preserves the intrinsic contact dissipation structure and guarantees physically consistent decay for general nonlinear systems.
    }
    \label{fig7}
\end{figure}

\subsection{Results}

\paragraph{Phase portrait.}

The contact variational integrator closely matches the RK4 reference solution in \(q(t)\) and \(p(t)\), confirming high trajectory accuracy.

\paragraph{Energy decay.}
Both methods exhibit the expected monotonic decay of
\begin{equation}\label{eq:H0-def}
    H_0(q,p) = \tfrac12\|p\|^2 + V(q).
\end{equation}
The theoretical dissipation law is
\begin{equation}\label{eq:dH0dt}
    \dot{H}_0 = -\gamma \|p\|^2,
\end{equation}
which the splitting method reproduces to high accuracy.

Figure \ref{fig7} depicts energy decay and convergence. The left panel compares energy decay for the exact potential law, energy splitting method, and RK4 scheme. The right panel shows a log-log plot confirming second-order convergence ($O(h^2)$), consistent with physically preserved dissipation by the integrator.

\paragraph{Conservative limit.}
When \(\gamma = 0\), the integrator becomes exactly symplectic and reduces to velocity-Verlet with no artificial numerical dissipation.

\begin{figure}[t]
    \centering
    \includegraphics[width=\linewidth]{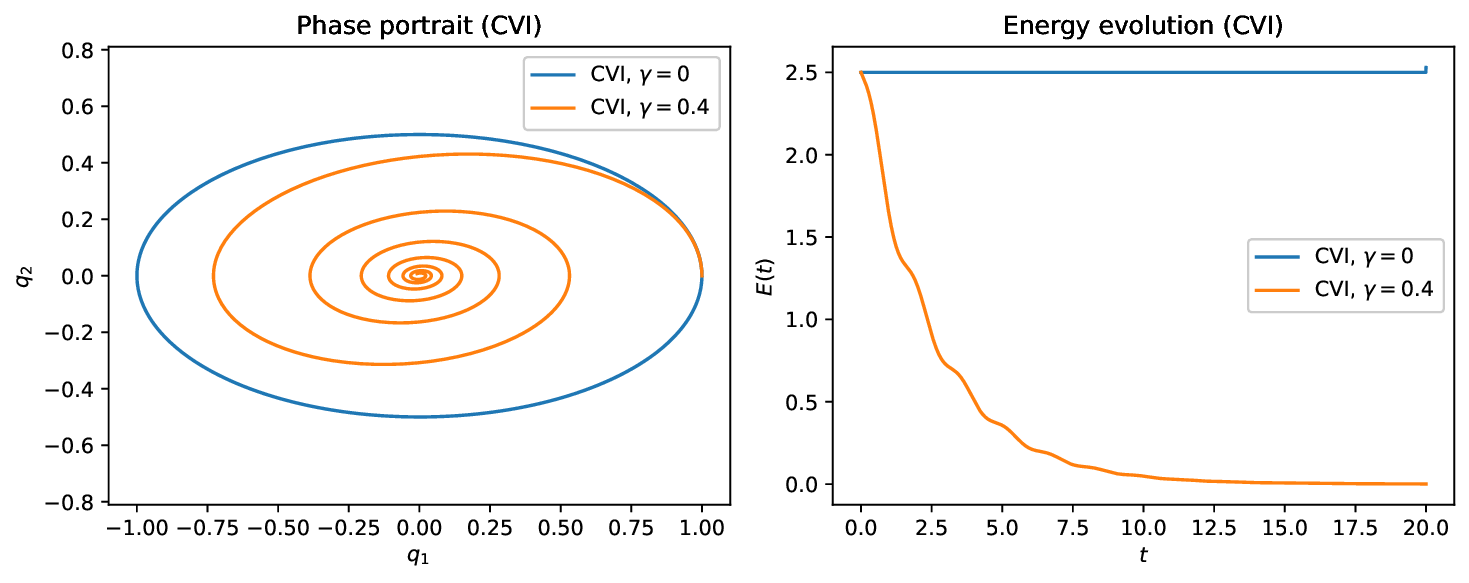}
    \caption{ Left: Phase portrait of the contact variational integrator (CVI) for the 2-DOF oscillator in the symplectic limit $\gamma=0$ and the dissipative regime $\gamma>0$.  
    For $\gamma=0$ the method reduces to a symplectic variational scheme and produces closed orbits, whereas for $\gamma>0$ the trajectories contract toward the origin, illustrating the built-in dissipative structure of the discrete Herglotz formulation.  
    Right: The corresponding energy evolution shows bounded oscillations in the symplectic case and monotone decay in the dissipative case, demonstrating that the CVI consistently captures both limits within a unified variational framework.
    }
    \label{fig8}
\end{figure}

Overall, the contact variational integrator provides a reliable, structure-aware method for simulating dissipative mechanical systems and recovers the standard symplectic variational integrator in the 
zero-damping limit.

Figure \ref{fig8} illustrates the symplectic versus dissipative limits of the contact variational integrator (CVI). The left panel compares phase space trajectories, while the right panel shows the corresponding energy evolution. In the absence of dissipation, trajectories remain closed (symplectic regime), whereas with dissipation, trajectories spiral toward the origin, demonstrating a smooth transition between conservative and dissipative dynamics.

\section{Conclusion}

We have presented a comprehensive geometric and computational framework for classical systems that exhibit both conservative and dissipative behavior. By formulating their dynamics on contact manifolds, we have shown how dissipation arises naturally through the Reeb direction, breaking conservation laws while preserving a generalized geometric structure. The derived momentum decay law and analysis of rigid-body dynamics on 
$SO(3)$ illustrate how contact geometry captures the interplay between symmetry, energy dissipation, and stability.

On the computational side, the contact variational integrator (CVI) developed here provides a robust and physically consistent method for simulating non-conservative dynamics. The scheme preserves the contact structure, reproduces the correct energy decay, and transitions smoothly to the symplectic Verlet integrator in the conservative limit. Numerical results confirm its second-order accuracy and faithful representation of irreversible dynamics.

Together, these results establish a unified bridge between geometric theory, physical interpretation, and numerical computation for dissipative mechanical systems. Future extensions may include stochastic contact systems, control applications, and thermodynamic generalizations, further connecting geometric mechanics with modern formulations of nonequilibrium dynamics.

\bibliographystyle{elsarticle-harv} 
\bibliography{geometry}
\bf{During the preparation of this work the author(s) used chatGPT in order to check spelling, grammar and presentation. After using this tool/service, the author(s) reviewed and edited the content as needed and take(s) full responsibility for the content of the published article.}
\end{document}